\documentclass[fleqn,10pt]{wlscirep}
\usepackage{braket}
\usepackage{siunitx} 

\title{Observation of rotational Brownian motion of single diamond nanoparticles}

\author[1,2*]{Masazumi Fujiwara}
\author[3,4,5]{Yutaka Shikano}
\author[2]{Ryuta Tsukahara}
\author[2]{Shinichi Shikata}
\author[2]{Hideki Hashimoto}
\affil[1]{Department of Chemistry, Osaka City University, Sumiyoshi-ku, Osaka, 558-8585, Japan}
\affil[2]{School of Science and Technology, Kwansei Gakuin University, Sanda, Hyogo, 669-1337, Japan }
\affil[3]{Institute for Molecular Science, National Institutes of Natural Sciences, Okazaki, Aichi, 444-8585, Japan}
\affil[4]{Institute for Quantum Studies, Chapman University, Orange, California 92866, USA}
\affil[5]{Research Center for Advanced Science and Technology (RCAST),
The University of Tokyo, Meguro-ku, Tokyo, 153-8904, Japan}
\affil[*]{masazumi@sci.osaka-cu.ac.jp}

\begin{abstract}
Capturing the rotational motion of single nanoparticles has been hindered owing to the difficulty of acquiring directional information under the optical diffraction limit. 
Here, we demonstrate that electron spins of single nitrogen vacancy (NV) centers can sense the rotational Brownian motion of their host nanodiamonds. 
When nanodiamonds are gradually detached from the substrates that they were fixed to, 
their optically detected spin resonance peaks are broadened by 1.8 MHz, 
which corresponds to the rotational diffusion constant of nanoparticles with a diameter of 11.4 nm from the Einstein--Smoluchowski relation.
\end{abstract}
\begin{document}

\flushbottom
\maketitle
\thispagestyle{empty}

\section*{Introduction}
Rotations of objects characterize one of the fundamental properties of the dynamics and can be measured by various optical techniques.
It is however significantly limited on the nanoscale owing to the difficulty of acquiring directional information under the optical diffraction limit. 
For characteristic dimensions above the sub-micron scale, optical microscopy provides ways to observe the rotational motion of small particles, 
such as capturing the anisotropic morphology of particles 
\cite{han2006brownian,cheng2003rotational,romodina2016detection},
polarization-sensitive optical detection of metallic nanorods~\cite{ruijgrok2011brownian,xiao2011imaging}, 
and extracting rotational information based on the detailed analysis of the 3D-translation dynamics~\cite{isojima2016direct}. 
On the other hand, at the atomic scale, fluorescence depolarization spectroscopy provides information 
on the rotational Brownian motion (diffusion) of ensembles of molecules~\cite{mann2003fluorescence}.
However, detecting rotational motion on the nanoscale, which lies between the atomic and the micron scales, 
has not been hitherto well explored.

The difficulty of detecting rotational motion on the nanoscale arises from two major technical challenges.
First, it is difficult to capture the shapes of nanoparticles by optical means due to their sizes being smaller than the optical diffraction limit.
Emission (either fluorescence or scattering) from nanoparticles is treated as coming from a point light source and cannot provide the directional information.
Second, the timescale of the rotational diffusion of nanoparticles is quite fast with a high dynamic range; for example, in water, 
it can vary from millisecond to microsecond for particle diameters from 100 nm to 10 nm, 
which is several orders of magnitude faster than that of micron-scale particles (1 \si{\um} gives 1.45 Hz)~\cite{berg1993random}.
Most of the image-based optical techniques cannot provide such high-frequency detection.
Thus, the detection of the rotational motion of nanoparticles has been elusive. 

A new approach that could access the rotational motion of single nanoparticles is to exploit the electron spins of 
nitrogen vacancy (NV) centers in nanodiamonds.
Nanodiamonds can be used as very stable fluorescnece nano-light emitters when incorporating NV centers~\cite{mochalin2012properties}.
The intensity of the NV fluorescence is electron-spin-dependent and can be affected by the nanoscale local environment of the nanodiamonds, 
such as magnetic fields, electric fields, and temperature~\cite{Doherty20131,schirhagl2014nitrogen}, which allows for quantum-enhanced nanoscale sensing~\cite{maze2008nanoscale,knowles2014observing,fujiwara2016manipulation}.
Applications of NV centers now extend to 3D-orientation tracking of nanoparticles~\cite{geiselmann2013three,Andrich2014,Horowitz2012} and nanoscale thermometry in living cells~\cite{kucsko2013nanometre,simpson2017non}. 
The electron spins of NV centers are in principle able to sense the rotational motion of the host nanodiamonds, 
because the random walk of the spin precession angle is accumulated as the geometric phase fluctuation of the NV quantum system.
The geometric phase fluctuation leads to dephasing of the electron spin coherence and thus broadens the electron spin resonance 
(ESR) line of NV centers in continuous-wave (CW) ESR detection~\cite{maclaurin2013nanoscale,maclaurin2010masterthesis,Ledbetter2012,Ajoy2012}.

Here, we report the linewidth broadening of the ESR lines of single NV centers by the rotational diffusion of the host nanodiamonds.
Single nanodiamonds incorporating single NV centers are slowly detached from the host substrates in an aqueous buffer solution.
Continuous optical measurement of the NV centers during the detachment process allows for measuring the ESR spectra of the same NV centers when the nanodiamonds are either fixed to the substrate or fluctuating during the detachment.
The ESR line is clearly broadened by 1.8 MHz (full width at half maximum, FWHM) by the nanodiamond fluctuation.
The observed broadening shows good agreement with the diffusion constant of nanodiamonds with a diameter of 11.4 nm, which is derived from the Einstein--Smoluchowski relation.
Our findings can provide a new method to measure the rotational motion of single nanoparticles and enable the exploration of nano-scale fluid mechanics.

\section*{Results}
\begin{figure}[t!]
\centering
\includegraphics[width=120mm]{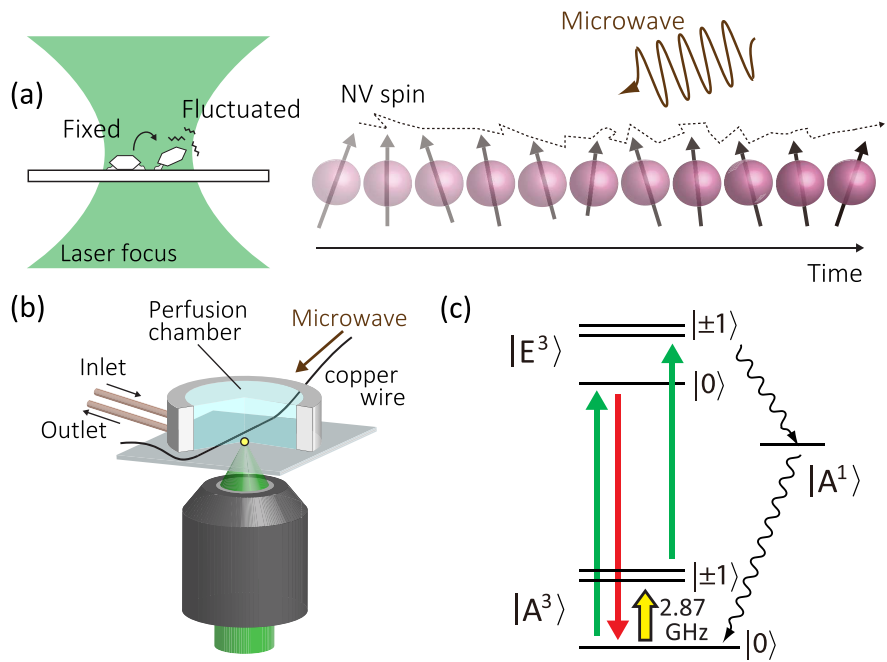}
\caption{(a) Schematic drawing of a single nanodiamond gradually detached from a coverslip it was attached to. The nanodiamond orientation (NV axis) is fluctuating due to the rotational Brownian motion. (b) Close up of the central part of the experimental setup. The nanodiamond is placed in a home-made perfusion chamber that simultaneously allows exchange of the solution and the optical experiment. A thin copper wire is fed through into the chamber for the spin excitation. Single NV centers hosted in nanodiamonds are excited by 532-nm laser light and are observed with red-shifted fluorescence collected through the same microscope objective. (c) Energy diagram of  NV centers. The main optical transition occurs between the ground state ($\ket{A^3}$) and the excited state ($\ket{E^3}$). Microwave excites the electron spin from $\ket{0}$ to $\ket{\pm1}$ in the sub-states of $\ket{A^3}$, followed by spin-conserving optical transitions and intersystem crossing from $\ket{\pm1}$ states in $\ket{E^3}$ to the lower singlet state,  ($\ket{A^1}$).  The population in $\ket{A^1}$ is nonradiatively relaxed to the spin $\ket{0}$ state of the triplet ground state, which causes a decrease of the fluorescence intensity.}
\label{fig1}
\end{figure}

\subsection*{Confocal fluorescence microscopy of singe NV centers in nanodiamonds in aqueous buffer solutions}

Figure~\ref{fig1}(a) shows a schematic drawing, 
which depicts nanodiamonds that are about to be detached from the coverslip.
We use commercially available nanodiamonds with a median size of 25 nm (Microdiamant, MSY0-0.05).
A droplet of the nanodiamond suspension is spin-coated on a coverslip.
The uniform nanodiamond distribution on the coverslip is confirmed by 
atomic force microscopy (see Supplementary Fig. S1).
We then fabricate a home-made perfusion chamber on the coverslip, 
which simultaneously allows liquid exchange and optical observation as 
shown in Fig.~\ref{fig1}(b) (see Methods and Supplementary Information for the details).
Distilled water is sent into the perfusion chamber to immerse nanodiamonds in liquid, followed by subsequent optical and ESR measurements. 
The total volume of the tube line and the perfusion chamber is 490 \si{\uL}.
The liquid flow keeps running with the flow rate of 80 $\si{\uL} \cdot \si{\min}^{-1}$ in the subsequent experiments 
unless specifically mentioned. 

\begin{figure}[b!]
\centering
\includegraphics[width=120mm]{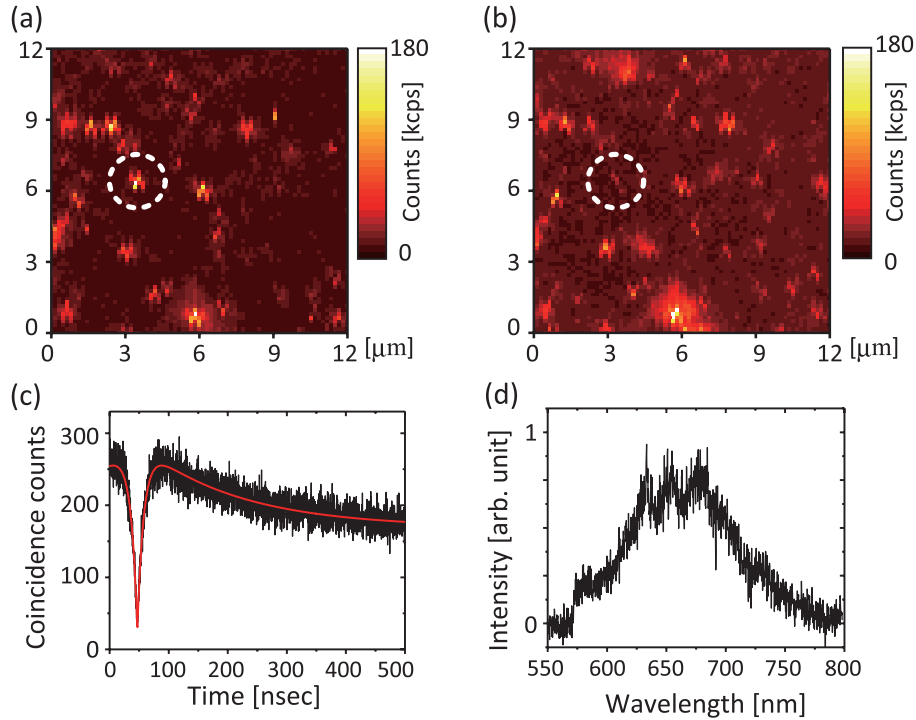}
\caption{(a) Confocal fluorescence scanning image of nanodiamonds deposited on a coverslip and immersed in water. The fluorescent nanodiamond indicated by the dashed circle is to be detached. (b) Scanning image of the same region after the nanodiamond is detached. (c) The second-order photon correlation histogram and (d) fluorescence spectrum of the nanodiamond, when it is fixed on the coverslip in water. }
\label{fig2}
\end{figure}

Figure~\ref{fig2}(a) shows a confocal fluorescence scanning image of nanodiamonds fixed on a coverslip under the flow of distilled water.
There are isolated nanodiamonds showing fluorescence signals, most of which are ascribed to NV centers.
Figures~\ref{fig2}(c) and (d) show the second-order photon correlation histogram and  fluorescence spectrum of the nanodiamond indicated by the dashed circle, respectively.
The photon-correlation histogram shows an antibunching dip with $g^{(2)}(0) = 0.16$  (the time origin $t_0 = 47$ ns) and a bunching shoulder at $t \sim 100$ ns due to the population exchange with the nearby metastable state~\cite{berthel2015photophysics} (see Fig.~\ref{fig1}(c)), 
which clearly indicates incorporation of a single negatively charged NV$^{-}$ center.
The temporal profile of the photon correlation data can be fitted with an equation reported elsewhere\cite{berthel2015photophysics}, 
which yields $\tau_1 = 11$ ns and $\tau_2 = 166$ ns.
The fluorescence spectrum is another clear signature of the presence of NV$^{-}$ centers; 
the zero-phonon line is observed at 634 nm, accompanied by a broad phonon sideband up to 750 nm
\cite{zhao2012effect,zhao2012suppression}.

We then replace the water with a buffer solution of pH = 7.5 and measure the ESR spectra.
The pH of the solution is changed stepwise (7.5 $\to$ 8.2 $\to$ 9.1 $\to$ 9.9) by adding a carbonate buffer solution.
After pH = 9.9 is reached, it is brought back to pH = 9.1 by adding HCl. 
A single process of changing  $\Delta {\rm pH} \sim 1$ takes about an hour that includes 
adjusting the pH of the reservoir, liquid circulation, and optical--ESR characterization.
The total time to change from water to pH = 9.1 (the final pH in this experiment) is about 6 hours.
When the pH is changed back to pH = 9.1, the fluorescence from the nanodiamond starts to fluctuate, 
as the nanodiamond is about to detach from the coverslip.  
During the time of the nanodiamond fluctuation of several minutes, we are able to measure the ESR spectrum, which will be described in the next section.
The nanodiamond finally moves away due to the continuous flow in the perfusion chamber.

Figure~\ref{fig2}(b) shows a confocal scanning image of the same region as imaged in Fig.~\ref{fig2}(a) 
after the nanodiamond is completely detached.
There is no more prominent fluorescent spot remaining inside the dashed circle, 
while other fluorescent nanodiamonds are still located at the same positions.
There remain residues that show very small fluorescence of 32 kcps; this is about 5 times smaller than the average fluorescence count of single NV centers detected in water in our laboratory setup ($\sim$150 kcps).
We therefore confirm that only the nanodiamond indicated by the dashed circle is removed during the pH change. Note that there are some new fluorescent blurred spots emerged in Fig.~\ref{fig2}(b) 
(see Supplementary Fig. S2 for the details).
These spots were created by the green laser excitation when we stopped the liquid flow during the experiment. 
Without the continuous liquid flow, the green laser excitation gradually generates such fluorescent spots 
(the fluorescence eventually grows much brighter than the single NV fluorescence if the liquid flow is stopped for a long time). 
This is probably because nanodiamonds detached from other locations (beyond the imaging region) 
are accumulated around the laser spot due to the strong green laser excitation (optical forces, laser heating, etc.) ~\cite{nishimura2014control}.
This phenomenon is more prominent in more acidic pH buffer solutions, which may be related to the zeta potential of nanodiamonds, 
as nanodiamonds show lower negative zeta-potentials for more acidic pH~\cite{williams2010size,petit2015probing,gines2017positive}.

\subsection*{ESR measurements on single NV centers}
\begin{figure}[b!]
\centering
\includegraphics[width=70mm]{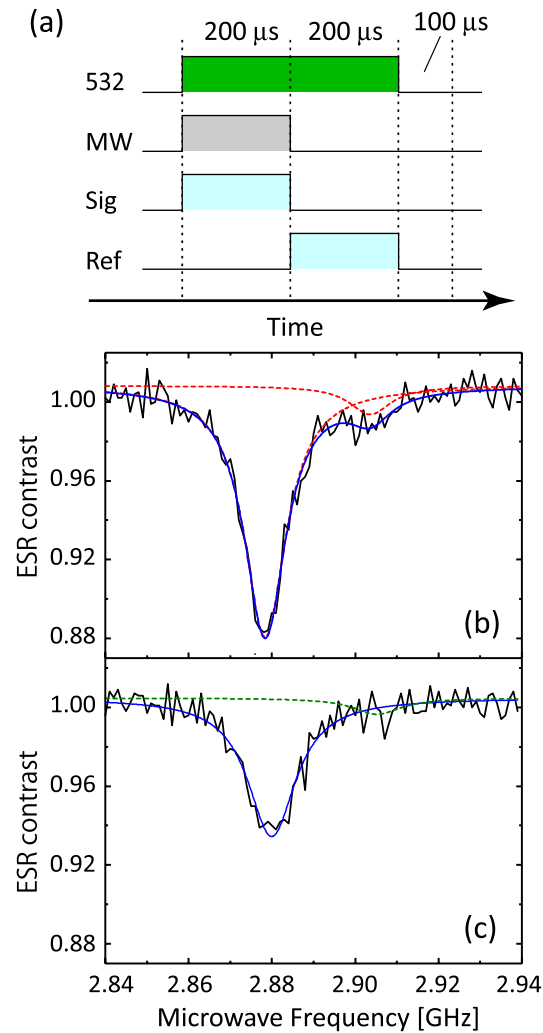}
\caption{(a) Schematic illustration of the gated photon counting for the ESR measurements. The APD detection is gated for microwave irradiation ON and OFF. 
The APD gate width is 200 \si{\us} common to the both gates, 
giving $\Delta I_{\rm PL} = I_{\rm PL}^{\rm ON} / I_{\rm PL}^{\rm OFF}$. 
The repetition rate of the gating (including laser off time of 100 $\si{\us}$) is 2 kHz.
The 532-nm green laser irradiation is continuously on during these gating periods. 
532: 532-nm green laser pulse. MW: microwave pulse. Sig: photon counting while the microwave is ON. 
Ref: photon counting while the microwave is OFF.
(b) ESR spectra of the nanodiamond indicated by the dashed circle (Fig.~\ref{fig2}) when it is fixed on the coverslip and (c) is fluctuating. The blue solid lines in (b) and (c) are the Lorentzian fits to the data, and the red dashed lines in (b) are the 2-peak components of the fitting. The olive dashed line in (c) is the reproduced curve for the minor peak, which is almost buried in the noise, 
calculated by taking account of the reduction of the main peak area.}
\label{fig3}
\end{figure}

We measure ESR spectra of the nanodiamonds throughout the course of the pH change. 
Figure~\ref{fig3}(a) shows a pulse sequence used to obtain the ESR spectra of single NV centers in nanodiamonds.
The microwave excitation and APD detection are gated with a common gate width of 200 $\si{\us}$ and a repetition rate of 2 kHz, 
in order to extract spin-dependent signals out of the fluorescence fluctuation 
due to the environmental noise such as defocusing of the laser spot, heating by the microwave irradiation, and  the nanodiamond fluctuation (see Methods)~\cite{Horowitz2012}.
The fluorescence intensities with/without the microwave excitation ($ I_{\rm PL}^{\rm ON}$ / $I_{\rm PL}^{\rm OFF}$) are measured, and their ratio $\Delta I_{\rm PL} = I_{\rm PL}^{\rm ON} / I_{\rm PL}^{\rm OFF}$ (ESR contrast) is plotted as a function of the microwave frequency.
The external magnetic field is not applied in this experiment.

Figure~\ref{fig3}(b) shows the ESR spectrum of a single NV center when the nanodiamond is fixed to the coverslip 
in distilled water (Fig.~\ref{fig2}(a)). 
The ESR spectrum is composed of two Lorentzian peaks.
Curve fitting with a two-peak Lorentzian profile determines that 
the major fluorescence peak is located at 2.8784(1) GHz with a linewidth of 12.2(4) MHz (FWHM), and 
another associated minor peak is located at 2.9036(11) GHz with a linewidth of 12.1(40) MHz (FWHM).
These two peaks are the result of intrinsic splitting of the magnetic sublevels of $\ket{\pm 1}$ spin states by lattice strain 
in the nanodiamond, 
which is often observed in nanodiamond NV centers~\cite{tisler2009fluorescence}. 
We note that the presented errors are from the curve fitting.

Figure~\ref{fig3}(c) shows the ESR spectrum of the single NV center while the fluorescence from the nanodiamond 
is fluctuating in the final buffer solution at pH = 9.1.
The major peak is clearly broadened, and the other minor peak is weakened.
We fit the data with a single Lorentzian profile, since the intensity of the minor associated peak is comparable to the noise level, 
which makes curve fitting difficult (the peak still exists at around 2.905 GHz). 
The major peak is located at 2.8800(3) GHz with a FWHM linewidth of 14.0(9) MHz.
The linewidth is broadened by 1.8(9) MHz compared with that of the fixed configuration (Fig.~\ref{fig3}(b)).
We note that the peak area of the major peak is decreased by 36.5 \% compared to that shown in Fig.~\ref{fig3}(b).
This decrease buries the minor peak under the noise, which justifies the use of a single Lorentzian profile for the curve fitting (we discuss this issue in Discussion).

\subsection*{Broadening of the ESR peak linewidth}
The observed linewidth  broadening comes from the rotational Brownian motion of the host nanodiamonds. 
Fast rotation of nanodiamonds adds a geometric phase to the time evolution of the NV spin system 
as theoretically studied in Refs.~\citenum{maclaurin2010masterthesis,maclaurin2013nanoscale}.
When the spin measurement time is sufficiently long compared with the rotational diffusion constant of the nanodiamonds, 
the final ESR signal is averaged over the entire ensemble of initial orientations and rotational trajectories. 
The random fluctuation of the geometric phase creates an additional decay channel for the quantum superposition, 
which modifies the spin coherence time as $T_2^\ast \to T_2^\ast + k_d^{-1}$.
Here, $k_d$ is the rotational diffusion constant of nanodiamonds and, according to the Einstein-–Smoluchowski relation, is given by 
\begin{equation}
k_d \left( =  \frac{\Delta \Gamma}{2} \right) = \frac{k_B T}{8 \pi (d/2)^3 \eta}, 
\end{equation}
where  $\Delta \Gamma$, $k_B$, $T$, $d$, and $\eta$ are the observed linewidth broadening in the ESR spectrum, 
Boltzmann constant, temperature, diameter of the nanoparticles, and viscosity of the surrounding medium, respectively.
Note that $\Delta \Gamma = 2k_d$, considering the time--frequency transformation relation of 
the exponential decay and the Lorentzian line shape.

In the present case, the surrounding medium is a carbonate buffer-based solution 
that has a viscosity of  $\eta = 0.97$  m\si{\Pa} at room temperature (T = 293 K) determined by a viscotester (Toki Sangyo, RE100L). 
The spin measurement time is 200 \si{\us}, sufficiently longer than the rotational diffusion time of 
the nanodiamonds, as it corresponds to the diffusion constant of nanoparticles with a diameter of 66 nm (5 kHz).
The observed linewidth broadening is hence calculated to be the rotational diffusion constant of 
nanoparticles with a diameter of $d$ = 11.4(2) \si{\nm}.
The size statistics of our nanodiamonds exhibit a mean diameter of 30 nm with a distribution ranging from 10 to 50 nm 
on the basis of the AFM topography image (Supplementary Fig.~S1).
The observed linewidth is within the range of the particle distribution, 
thus indicating that it comes from the rotational Brownian motion.

\section*{Discussion}
In this paper, we reported the linewidth broadening of the ESR peaks of single NV centers 
due to the rotational Brownian motion of the host nanodiamonds. 
We found that the ESR peak was broadened by 1.8 MHz (FWHM) that corresponds to the rotational diffusion constant of the nanodiamonds with a diameter of 11.4 nm.
While the results clearly demonstrate the effect of the rotational Brownian motion on the ESR peaks of NV centers, 
there still remain questions that need to be addressed to fully understand the presented results. 

\subsubsection*{Other possible factors contributing to the linewidth broadening} 
Laser heating of the nanodiamonds by the 532-nm green laser may affect the linewidth broadening as it can change the local viscosity.
The local temperature in the present situation can rise up to, for example, 90 $^\circ {\rm C}$ 
(slightly lower than the boiling point of water), which reduces the viscosity of water to almost one third of the original value~\cite{korson1969viscosity} 
(1.0 m\si{\Pa} at 20 $^\circ {\rm C}$ $\to$ 0.31 m\si{\Pa} at 90  $^\circ {\rm C}$). 
The change of the local viscosity may lead to overestimation of the linewidth broadening by a factor of 3, giving the 
corresponding particle size of $d$ = 16.4(3) \si{\nm}.
Such temperature change, however, might not occur in the present situation, 
since we did not observe the peak shift that corresponds to this temperature change of -0.7 MHz (the temperature change causes the peak shift of -74 kHz$\cdot {\rm K}^{-1}$~\cite{acosta2010temperature}).
Note that the microwave power and the laser intensity can affect the linewidth, which we describe in the following subsection.

The pH change does not cause such a drastic change of the linewidth as observed here. We have investigated the effect of the pH on the electron spin properties of single NV centers in the same nanodiamond samples. 
The variation of the linewidth over the pH change from 4 to 11 is always smaller than 0.6 MHz~\cite{tracking2018fujiwara}.

It is interesting to note that the linewidth broadening of the ESR peaks has been observed in optically trapped nanodiamonds with a relatively large diameter of 74 nm~\cite{geiselmann2013three}. 
The nanodiamonds used in Ref.~\citenum{geiselmann2013three} have a mean particle size of 74 nm based on the dynamic light scattering  data 
and are trapped in a viscous solution (glycerin:water = 5:1, viscosity 132 m\si{\Pa} at 20 $^\circ {\rm C}$~\cite{cheng2008formula}),  
which gives $k_D$ of $\sim$ 24 Hz. 
A detectable broadening on the kHz level was confirmed by statistically comparing the ESR spectra of the trapped nanodiamonds with those of the fixed ones (Fig. S9 in Ref.~\citenum{geiselmann2013three}) and qualitatively ascribed to the precession of the NV axis. 
Further quantitative evaluation would be important to reveal the detailed mechanism of the linewidth broadening.

\subsubsection*{Reduced peak area during the fluctuation}
The peak area of the major peak in Fig.~\ref{fig3}(c) is decreased by 36.5 \% compared to that in Fig.~\ref{fig3}(b).
This peak reduction results in the associated minor peak being buried under the noise level, 
which makes it impossible to fit with the original double Lorentzian peak profile. 
By applying the same change (36.5\%-peak-area reduction and 1.6-MHz-frequency shift) to the fitting parameters of the minor peak in Fig.~\ref{fig3}(b) and keeping other parameters fixed (see Supplementary Table~S1), we reproduce the simulated curve for the minor peak as shown in Fig.~\ref{fig3}(c) (dashed olive line).
The reproduced minor peak is smaller than the noise, 
which justifies the use of a single Lorentzian profile for the curve fitting. 

Both the microwave power and the laser excitation power can affect the peak area, 
as the ESR linewidth exhibits corresponding dependences~\cite{Lesik2011}. 
We measured the dependence of the ESR peaks on both the microwave power and 
laser power for nanodiamonds fixed to a coverslip in water (see Supplementary Fig. S3) to
estimate the effect of these parameters on the reduction of the peak area (and linewidth broadening).
In the present situation, the linewidth can be broadened by $\pm$ 0.2 MHz by a typical microwave power change of 1 \% and 
$\pm$ 0.2 MHz by a typical laser intensity fluctuation of 5--10 \%. 
The observed linewidth broadening and the peak area reduction therefore do not originate from the power change of the 
microwave and the laser, confirming the origin of the linewidth broadening as the rotation Brownian motion of the nanodiamonds.

\subsubsection*{Future perspectives of rotational-motion sensing}
Despite of these unresolved issues, 
the present observation suggests attractive applications of this rotational-motion quantum sensing to various fields like nanofluidics or 
biological sensing.
On the nanoscale, the persistent photostability of NV centers, together with the present rotational-motion sensing, will make nanodiamonds indispensable tools to investigate nanoscale fluid mechanics.
The classical experimental tool to visualize nanofluids is organic-molecular fluorescent probes~\cite{fornander2016visualizing,sinton2004microscale} that can also be used for fluorescence depolarization spectroscopy~\cite{smith2015review}.
This method, however, suffers from bleaching of dyes and can only be used for a short period of time and in some specific conditions at specific pH and temperature range. 
Furthermore, the nanoscale volume restricts the number of fluorescent molecules, thereby shortening the observation time further.
In contrast, fluorescent nanodiamonds can provide long-term tracking in various pHs and temperature ranges with excellent fluorescence stability.
Recent advancement of fabrication technology has enabled the incorporation of NV centers in nanodiamonds smaller than 5 nm~\cite{bradac2010observation,tisler2009fluorescence}.
One could insert such ultra-small nanodiamonds into structures with a size of tens of nanometers.

It is also possible to access the translational Brownian motion at the same time with measuring the rotational Brownian motion 
by a wide-field imaging technique. 
Combining NV-fluorescent nanodiamonds with walking protein motors would be interesting because NV centers provide 
information on the 3D protein motion (rotation or torsion in addition to the translational motion)~\cite{isojima2016direct}.
Our NV sensing technique can thus provide a way for extracting full information on the Brownian motion of protein motors in fluorescence microscopes.

\section*{Methods}
\subsection*{Sample preparation}
A commercially available nanodiamond suspension (Microdiamant, MSY 0-0.05, median particle size: 25 nm) was purified by centrifugation and was dispersed in pure water.
A small droplet of the suspension was spin-coated on a cleaned coverslip. 
A 25-\si{\um}-thin copper wire was placed on the coverslip as a microwave linear antenna, and both ends were soldered to SMA connectors.
An acrylic spacer with a height of about 4 mm with inlet and outlet tubes was then glued 
on top of the sample using  a UV-curing resin. 
It was sealed with a glass plate to make a perfusion chamber.
The spin-coated samples were raster scanned with an atomic force microscope (Bruker, Edge) to obtain topography images. 
The peak heights of the distributed nanodiamonds were measured to extract the particle size distributions. 
The nanodiamonds were detached from the substrate by changing the pH of the buffer solution stepwise in the perfusion chamber.
We first sent distilled water to the chamber.
A flow of pH-buffer solution (0.1 M sodium carbonate with 5\% HCl) was next created.
During the optical excitation, the continuous flow of these solutions with a rate of 80 $\si{\uL} \cdot {\rm min}^{-1}$ was maintained to prevent photothermal accumulation of nanodiamonds.

\subsection*{Optical measurements}
The perfusion chamber was mounted on a 3-axis piezo stage in a home-built confocal fluorescence microscope.
A continuous-wave 532 nm laser was used for the excitation with an excitation intensity of 94 kW$\cdot {\rm cm ^{-2}}$ (250 \si{\uW}).
An oil-immersion microscope objective with a numerical aperture of 1.4 
was used both for the excitation and the fluorescence collection. 
The NV fluorescence was filtered by a dichroic beam splitter (Semrock, FF560-FDi01) 
and a long-pass filter (Semrock, BLP01-561R) to remove the residual green laser scattering. 
It was then coupled to an optical fiber acting as a pinhole (Thorlabs, 1550HP, core diameter $\sim 10$ \si{\um}).
The fiber-coupled fluorescence was finally guided into a Hanbury-Brown-Twiss (HBT) setup consisting of two APDs (Perkin Elmer SPCM AQRH-14) and a 50:50 beam splitter or connected to a spectrometer equipped with a liquid-nitrogen-cooled CCD camera (Princeton, LNCCD). 
By scanning the sample with the piezo stage, we were able to obtain the fluorescence scanning images.
A time-correlated single-photon counting module (PicoQuant, TimeHarp-260) was used to obtain second-order photon correlation histograms.

\subsection*{ESR measurements}
Microwave generated from a source (Rohde \& Schwarz, SMB100A) was boosted by 45 dB with an amplifier (Mini-circuit, ZHL-16W-43+) and was fed to the microwave linear antenna in the perfusion chamber.
The microwave excitation power was 35 dBm (3.2 W).
To extract the ESR spectra from the fluorescence fluctuation of the nanodiamonds, 
the APD detection was gated for microwave irradiation ON and OFF states
by using an RF switch (Mini-circuit, ZYSWA-2-50DR-S) and a bit pattern generator (Spincore, PBESR-PRO-300). 
The gate width was 200 \si{\us} common for both gates, followed by a laser shut-off time of 100 $\si{\us}$, 
giving $I_{PL}^{ON}$ and $I_{PL}^{OFF}$. 
The repetition rate of the gating pulses was 2 kHz.
No external magnetic field was applied.


\section*{Acknowledgements}
We thank Prof. Noboru Ohtani for the AFM measurements.
MF acknowledges the financial supports from JSPS-KAKENHI (Nos. 26706007, 26610077, 16K13646, and 17H02741), MEXT-LEADER program, and Osaka City University (OCU-Strategic Research Grant 2017 for young researchers).
YS acknowledges the financial support from JST ERATO (Grant No. JPMJER1601). 
SS acknowledges the financial support from JSPS-KAKENHI (No. 26220602).
HH thanks JSPS KAKENHI, Grant-in-Aids for Basic Research (B) (No. 16H04181) and Scientific Research on Innovative Areas "Innovations for Light-Energy Conversion (I$^4$LEC)" (Nos. 17H06433, 17H0637) for financial support

\section*{Author contributions statement}
MF designed the research, performed the experiment, and analyzed the data. YS analyzed the data. RT performed sample characterization. All the authors discussed and wrote the paper.

\section*{Additional information}
\textbf{Competing financial interests}: The authors declare no competing financial interests. 

\end{document}